\documentclass[acmsmall]{acmart}

\usepackage{float}
\usepackage{subfigure}
\AtBeginDocument{%
  \providecommand\BibTeX{{%
    \normalfont B\kern-0.5em{\scshape i\kern-0.25em b}\kern-0.8em\TeX}}}
\usepackage{tabularx}
\usepackage{bbding}
\usepackage{graphicx}
\usepackage{geometry}
\usepackage{subfigure}
\usepackage{amsmath}
\usepackage{makecell}
\usepackage{float}
\usepackage{color}
\usepackage{booktabs}
\usepackage{caption}
\usepackage{tabu}
\usepackage{hyperref}
\usepackage{csquotes}
\usepackage{xcolor}

\copyrightyear{2025}
\acmYear{2025}
\setcopyright{acmlicensed}\acmConference[CSCW '25]{Proceedings of the 2025 ACM Conference on Computer Supported Cooperative Work}{October 18--22, 2025}{Bergen, Norway}
\acmBooktitle{Proceedings of the 2025 ACM Conference on Computer Supported Cooperative Work (CSCW '25), October 18--22, 2025, Bergen, Norway}
\acmPrice{15.00}
\acmDOI{}
\acmISBN{}

\newcommand{\add}[1]{{\color{blue} #1}}

 \author{Zhiting He}
 \email{zhiting@cuc.edu.cn}
 \orcid{0009-0005-3906-1681}
 \affiliation{
 \institution{Communication University of China}
 \city{Beijing}
 \country{China}}

  \author{Jiayi Su}
 \email{etnentryno.1@gmail.com}
 \affiliation{
 \institution{Guangzhou Academy of Fine Arts}
 \city{Guangzhou}
 \country{China}}

   \author{Li Chen}
 \email{lchen488-c@my.cityu.edu.hk}
 \affiliation{
 \institution{City University of Hong Kong}
 \city{Hong Kong}
 \country{China}}

 \author{Tianqi Wang}
 \email{tttttiq@outlook.com}
 \affiliation{
 \institution{Guangdong University of Technology}
 \city{Guangzhou}
 \country{China}}

 \author{RAY LC}
 \authornote{Correspondences should be addressed to LC@raylc.org}
 \email{LC@raylc.org}
 \orcid{0000-0001-7310-8790}
 \affiliation{
 \institution{City University of Hong Kong}
 \city{Hong Kong SAR}
 \country{China}}

\begin{document}
\title[I Recall the Past]{"I Recall the Past": Exploring How People Collaborate with Generative AI to Create Cultural Heritage Narratives}

\begin{abstract}
Visitors to cultural heritage sites often encounter official information, while local people's unofficial stories remain invisible. To explore expression of local narratives, we conducted a workshop with 20 participants utilizing Generative AI (GenAI) to support visual narratives, asking them to use Stable Diffusion to create images of familiar cultural heritage sites, as well as images of unfamiliar ones for comparison. The results revealed three narrative strategies and highlighted GenAI's strengths in illuminating, amplifying, and reinterpreting personal narratives. However, GenAI showed limitations in meeting detailed requirements, portraying cultural features, and avoiding bias, which were particularly pronounced with unfamiliar sites due to participants' lack of local knowledge. To address these challenges, we recommend providing detailed explanations, prompt engineering, and fine-tuning AI models to reduce uncertainties, using objective references to mitigate inaccuracies from participants' inability to recognize errors or misconceptions, and curating datasets to train AI models capable of accurately portraying cultural features.
\end{abstract}




\maketitle

\section{Introduction}\label{sec:Introduction}

People visit cultural heritage sites for diverse reasons, engaging in a meaningful dialogue with the history, culture, and humanities intertwined with them. The tangible and intangible elements of cultural heritage that they can see, hear, and touch collectively shape a unique cultural ambiance, offering valuable experiences and knowledge. Within these experiences and knowledge, the social values linked to the historic environment stand out as a crucial characteristic of cultural heritage ~\cite{jones_wrestling_2017}. This encompasses not only historical value as defined by heritage professionals but also includes memory, oral history, symbolism, and cultural practices associated with the historic environment ~\cite{jones_wrestling_2017}. However, visitors to cultural heritage sites are often presented with information related to heritage history, significant events, and important status provided by official historical institutions, known as Authorized Heritage Discourse (AHD) ~\cite{tsenova_-authorised_2020}. In contrast, unofficial memories, emotions, and stories related to local people are often transient and invisible. Therefore, heritage studies underscore the importance of presenting the valuable perspectives and stories of communities and individuals familiar with the place, rather than relying solely on official narratives ~\cite{han_enhancing_2014,tsenova_loci_2023,lu_participatory_2023}. 

However, the challenge in collecting community stories is that, although individuals familiar with heritage sites may possess expertise regarding their cultural heritage ~\cite{clarke_familiar_2018,clarke_repeat_2021}, they are not necessarily narrative experts and lack the means and tools to participate in familiar cultural heritage storytelling. Therefore, we aim to explore the potential of generative AI technology in promoting individuals' participatory activities in the contexts of culture heritage ~\cite{giglitto2019bridging} The openness and open-source nature of some generative AI tools ~\cite{chiou_designing_2023, walsh_ai_2019}, along with their auxiliary applications in the creative field, make these AI tools have the potential to lower the barriers to narrative creation, enabling more people to access cultural narratives. This, in turn, facilitates sustained personal storytelling and amplifies diverse voices within the cultural narrative space ~\cite{tsenova_-authorised_2020}. Moreover, considering that the data supplied to the AI encompasses a diverse range of perspectives, it can influence individuals' narratives and provide a distinctive visual perspective throughout the generation process, thus fostering inspiration and reflection ~\cite{chiou_designing_2023}. However, generative AI also comes with limitations, such as the potential for cultural bias that may mislead individuals unfamiliar with the cultural heritage. Therefore, we aim to explore the ways, advantages, and challenges of using generative AI to assist individuals in cultural narrative creation, and propose possible solutions.

Therefore, the research questions that we aim to answer are: 

\textbf{RQ1}: How do participants use narrative strategies when collaboratively creating narratives about familiar cultural heritage with generative AI?

\textbf{RQ2}: What are the strengths and limitations of generative AI in supporting the creation of personal narratives about familiar cultural heritage?

\textbf{RQ3}: How does the creation of personal narratives about unfamiliar cultural heritage, without prior knowledge and experience, differ from that of narratives related to familiar cultural heritage?

To pursue these research objectives, we conducted workshops with 20 participants. During the workshops, participants used Stable Diffusion to generate a series of images related to familiar cultural heritages, such as those from their hometowns or locations where they spent considerable time ~\cite{pearce_experience_2012}, to create a cultural narrative. In subsequent semi-structured interviews, participants shared their creative themes, insights, and challenges encountered during the collaborative creation process with AI. We also asked them to create narratives about unfamiliar cultural heritages to compare and explore how the level of familiarity with heritage sites influenced their creations. 

Our results indicated that (1) participants adopted three distinct narrative strategies, ranging from wholly depicting objective heritage sites to entirely expressing subjective interpretations, with some participants changing their strategies midway through the process; (2) AI assisted in illuminating, amplifying, and reinterpreting personal narrative creation; (3) the cultural limitations, cultural bias and uncertainty of AI had adverse effects on personal narrative; and (4) participants were more prone to overlooking mistakes and biases within output images about unfamiliar cultural heritage. These findings shed light on potential methods, strengths, and limitations in using generative AI to create personal narratives of familiar cultural heritage, thereby demonstrating its potential to facilitate individuals' access to cultural narratives.

\section{Related Work}\label{sec:Background}

\subsection{Cultural Heritage Narrative}

Cultural heritage narratives can enhance individuals' comprehension and documentation of cultural heritage ~\cite{giglitto_enhancing_2015,muntean_designing_2017,liu_digital_2023,gibson_its_2023,lu_shadowstory_2011,schaper_enhancing_2022}. Personal narratives about cultural heritage have the capacity to promote social cohesion by addressing the marginalization and undervaluation of public participation in heritage archives ~\cite{giglitto_enhancing_2015,tsenova_loci_2023,lu_participatory_2023,bala_towards_2023}. 
These individual accounts can further enrich the understanding and perception of cultural heritage for others ~\cite{muntean_designing_2017,giaccardi_heritage_2012}, fostering a more profound resonance within the community and broader contexts ~\cite{nisi_before_2023,claisse_crafting_2020,tsenova_-authorised_2020,zhao_involving_2023,alelis_exhibiting_2013,allen_interactive_2013}. 

A familiar place refers to individuals' hometowns as well as locations where they have spent considerable time and consequently feel familiar and have extensive previous knowledge of the setting~\cite{pearce_experience_2012}. Familiar places are highly valued in familiar tourism. For example, Clarke and Bowen's investigation into the behaviors and attachments of familiar tourists claimed that locals have extensive knowledge of their area, enabling them to participate in tourism co-creation and become valuable providers of tourism resources~\cite{clarke_familiar_2018}. They also mentioned that the community has a historical, storytelling perspective on familiar places~\cite{clarke_repeat_2021}. Hung et al. applied VR to enable people to revisit well-known places, illustrating how modern technology can expand and enrich the cultural experience of familiar tourists~\cite{hung_exploring_2024}. These studies reveal the valuable knowledge and understanding that familiar visitors have of places. We believe that locals' familiarity with their area also encompasses an understanding of local cultural heritage, which can provide valuable individual perspectives to cultural narratives.

Narrative is defined as a structured form that people use to organize stories, events, and facts related to their experiences, memories, or understanding of the world, thereby rendering these elements meaningful~\cite{malegiannaki_teaching_2020}. Empowering familiar groups to narrate their cultural heritage stories leverages this potential of narratives. Tsenova et al. defined narrative as encompassing multiple stories, offering a broad perspective through which people make sense of their surroundings, with each story focusing on individual characters or personal experiences~\cite{tsenova_loci_2023}. Nevertheless, the challenge is evident when individuals residing in heritage sites, despite possessing knowledge about their cultural heritage and personal experiences, lack proficiency in narrative techniques. Consequently, they might encounter difficulties in utilizing tools to effectively convey their stories. Therefore, it becomes imperative to minimize entry barriers, facilitating sustained personal storytelling and thereby amplifying a diverse range of voices within this domain~\cite{tsenova_-authorised_2020}.

Recent studies underscore the need for employing technology to facilitate participatory activities that enhance community engagement with cultural heritage. Do et al. demonstrated this through a digital photo display system that actively involved a rural community in sharing and preserving their local heritage ~\cite{do_content_2015}. Similarly, Echavarria et al. engaged younger demographics by using augmented reality to transform local heritage sites into interactive learning environments ~\cite{echavarria_creative_2022}. Giglitto et al. highlighted the power of digital tools in enabling marginalized groups to participate more fully in cultural heritage activities, thereby promoting inclusivity ~\cite{giglitto2019bridging}. Ciolfi et al. further emphasized the need for collaborative platforms that bring together professionals and community members in heritage conservation efforts ~\cite{ciolfi_cultural_2015}. Collectively, these studies reflect a strong communal interest in leveraging technology to boost engagement and ensure broad participation in cultural heritage preservation.

Technological advancements significantly enhance how communities and individuals narrate cultural heritage. Kambunga et al. proposed that participatory design and interactive exhibitions allow for the excavation of personal narratives within cultural contexts, engaging communities in the preservation and dissemination of cultural heritage~\cite{kambunga_participatory_2020}. Similarly, the work by Giglitto et al. emphasizes the role of digital technologies in facilitating community participation in cultural heritage, particularly for marginalized groups such as migrants and refugees, enhancing social inclusion through tailored digital tools that respect cultural nuances~\cite{giglitto2019bridging}. Shin and Woo applied augmented reality (AR) storytelling at cultural heritage sites to create site-specific narratives that enhance the historical and cultural context for visitors, making the narrative experience immersive and localized~\cite{shinHowSpaceTold2023}. Nisi et al. claimed that using digital narrative tools facilitates the sharing of immigrant experiences and cultural heritage, promoting social inclusion and enhancing public dialogue through story-driven digital tools~\cite{nisi_before_2023}. AI technologies further support these endeavors by enabling sophisticated narrative analytics and personalized content generation, thus enriching the narrative landscape of cultural heritage~\cite{fu2024being,wu2024present}. Bernstein et al. proposed that generative AI can significantly enhance individual access to narrative creation by lowering the technical entry threshold, thus democratizing the tools necessary for cultural expression~\cite{bernstein_architecting_2023}. Similarly, Muller et al. claimed that AI technology enables even those without deep technical or creative training to craft complex, creative content, enriching cultural narratives with diverse perspectives~\cite{muller_genaichi_2024}. As a result, generative AI has the potential to allow a broader community to contribute their stories and experiences, thereby enriching the cultural narrative in profound ways.

In this study, we aim to involve individuals in collaborative efforts with generative AI to create personal narratives about cultural heritage, thus exploring the potential of generative AI in enhancing individual access to cultural narratives.

\subsection{Supporting Cultural Heritage Narratives with Generative AI}

Generative AI uses existing media as data to produce new media in various formats such as images, texts, and videos ~\cite{muller_genaichi_2024}. Generative AI has exerted influence on users in diverse domains ~\cite{suh_ai_2021,kamath_evaluating_2023,louie_expressive_2022,druga_family_2022,pataranutaporn_living_2023,yang_ai_2022}. often making creative tasks more accessible by democratizing the tools necessary for creation and engagement ~\cite{qiao_initial_2022,weisz_perfection_2021,weisz_better_2022}. In the realm of social media, Lyu et al. applied generative AI to reduce content creation costs and break down skill barriers, thereby augmenting human creativity and enabling broader participation in content generation, as evidenced in the use of generative tools by YouTubers~\cite{lyu_preliminary_2024}. Louie et al. applied these technologies to empower untrained users to engage in complex creative endeavors such as music composition, traditionally requiring specific skills and training~\cite{louie_novice-ai_2020}. Moreover, generative AI has shown promise in art and design ~\cite{caramiaux_explorers_2022,kim_colorbo_2022,lawton_drawing_2023,ling_sketchar_2024}, such as fostering 3D modeling ~\cite{liu_3dall-e_2023,zhang_generative_2023}, supporting creative artwork-making ~\cite{lc_imitations_2022,lc_human_2023,lc_together_2023}, and aiding in more interactive and inclusive design processes ~\cite{walsh_ai_2019,chiou_designing_2023,rezwana_understanding_2022,long_designing_2019,lc_speculative_2024}. Guo et al. applied AI tools like DALL-E and Stable Diffusion to facilitate the participation of individuals in artwork creation by lowering technical barriers and increasing productivity ~\cite{han_when_2024}, fostering broad community participation~\cite{guo_exploring_2024}. This extends to design, where research team developed the BrainFax platform integrated with communication tools to enhance collaborative design efforts, making visual creation accessible to designers irrespective of their visualization skills~\cite{verheijden_collaborative_2023}. Generative AI also plays a transformative role in cultural engagement. Trichopoulos used GPT-4 as an interactive guide in museums, enhancing the visitor experience through tailored narratives and recommendations based on user preferences, thus making cultural education more immersive and personalized~\cite{trichopoulos_large_2023}. These examples underscore how generative AI can affect community and cultural participation by making sophisticated tools accessible and fostering creative expression.

While generative AI unlocks tremendous potential in creativity and cultural engagement, it also presents significant challenges, particularly concerning ethics, accessibility, and user understanding. Verheijden and Funk claimed that technical limitations arise in visualizing complex or abstract prompts, revealing the constraints of AI in accurately rendering specific functionalities~\cite{verheijden_collaborative_2023}. They also emphasized that ethical concerns are prevalent, with issues like data provenance and the potential reproduction of biases in AI-generated images raising alarms about the integrity of AI models~\cite{verheijden_collaborative_2023}. Lyu et al. pointed out that accessibility challenges are highlighted by the ambiguity and limited explainability of AI systems, which can alienate non-expert users and erode trust in these technologies~\cite{lyu_preliminary_2024}. Guo et al. highlighted that the regulation of online creativity also faces threats from AI's capability to produce low-quality and potentially infringing content, triggering conflicts within creative communities~\cite{guo_exploring_2024}. Moreover, Muller et al. proposed that the unpredictable nature of AI-generated content complicates user control, questioning the effectiveness of current user interface paradigms and necessitating a reevaluation of interaction design~\cite{muller_genaichi_2024}. The shift from user-driven to model-driven control underscores the need for designing safer and more effective interactions tailored to generative AI~\cite{weisz_design_2024}. Mahdavi Goloujeh pointed out that ethical dilemmas extend to the sharing of prompts and creative ownership in community settings, where the balance between inspiration and originality must be carefully managed to avoid cultural and ethical missteps~\cite{mahdavi_goloujeh_social_2024}. In cultural contexts, Trichopoulos claimed that the deployment of large language models demands meticulous implementation and continuous evaluation to maintain cultural integrity and ensure inclusivity~\cite{trichopoulos_large_2023}. Addressing these multifaceted challenges is crucial for fostering an environment where generative AI can be both innovative and ethically sound, enhancing its benefits while mitigating risks.

Therefore, it is crucial to explore the potential of generative AI in enabling individuals to better participate in cultural narratives. AI technology holds the promise of enhancing human creativity through co-creation, as seen in design and music, where it supports the creation of high-quality outputs and empowers users with greater control and self-efficacy~\cite{verheijden_collaborative_2023}. Generative AI also facilitates richer engagement in cultural heritage, providing personalized and informative experiences that can transform visitor interactions~\cite{trichopoulos_large_2023}. However, AI also has limitations, including the need for careful integration to avoid overshadowing human creativity and ensuring that ethical standards and cultural integrity are maintained. By addressing these challenges, we can maximize the benefits of generative AI while providing a more inclusive and creative community engagement.

\section{Methods}\label{sec:Methods}

\subsection{Workshop Tasks and Procedure}
We conducted workshops to investigate human-AI co-creation of personal narratives related to cultural heritage. Each workshop was held virtually using an online meeting tool and lasted around 1.5 to 2 hours, involving one participant and one researcher. The workshop included a pre-study interview, a training session, two distinct image generation processes of familiar and unfamiliar cultural heritage. Each generation process was followed by a post-study semi-structured interview (see Fig.1). Stable Diffusion ~\cite{rombach2022high} was chosen as the image generation tool because, compared to other generative AI tools such as MidJourney, it is open-source, making it accessible to more people. Additionally, it offers more parameter options and models, allowing participants to flexibly manipulate the style and content of generated images to meet their requirements. We utilized version 1.3.2 of the local version of Stable Diffusion, along with the v5-PrtRE version of the Anything Model. All procedures followed institutional guidelines IRB for human subject testing.

\begin{figure}[htbp]
  \centering
  \includegraphics[width=1\textwidth]{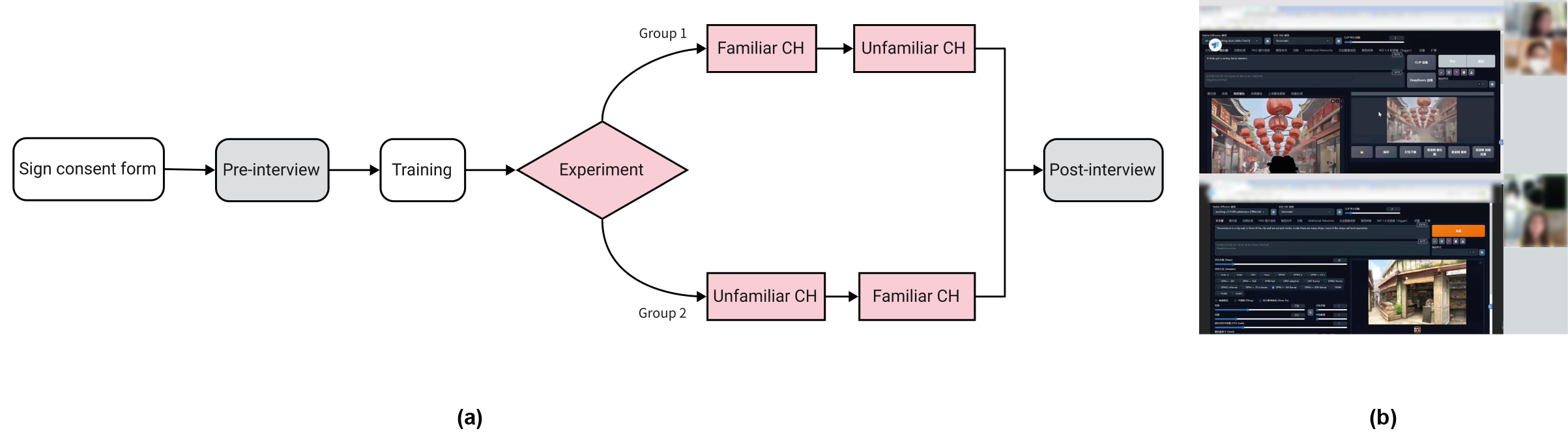}
  \caption{(a) is procedure diagram, (b) is the screenshot of the online workshop meeting.}
  \label{fig:UW-Workflow}
  \Description{Caption}
\end{figure}

\subsubsection{Introduction to the Task}

After signing a consent form, participants were given the opportunity to use Stable Diffusion to complete two image creation tasks: one related to familiar cultural heritage and another related to unfamiliar cultural heritage. For the familiar cultural heritage task, participants were asked to first briefly recall and describe their impressions and personal experiences with the familiar cultural heritage to evoke their memories, and then create an image about this familiar cultural heritage, with no restrictions on the theme. For the unfamiliar cultural heritage task, participants were instructed to choose one of the ten provided cultural heritage sites they were unfamiliar with. We showed them around three pictures and a brief introduction to give them a general idea of the place. After that, they were asked to create an image about this unfamiliar cultural heritage based on their own understanding and guesses, with no restrictions on the theme. They could continue generating images until they were satisfied with the result.

We chose Hong Kong as the unfamiliar cultural heritage site due to our familiarity with the locations, which allowed us to identify overlooked details in the images generated by participants. Additionally, we presented 10 distinct cultural heritage sites for participants to choose from, ensuring that each site was selected almost equally to minimize differences in their influence on the generated results, thereby avoiding any bias resulting from specific site characteristics.

\subsubsection{Specific Process}
Participants remotely controlled our computer to generate images, ensuring uniformity in equipment usage among all participants. Following the introduction, we presented a video tutorial illustrating the utilization of three features in Stable Diffusion for image generation and adjustment—specifically, text-to-image and inpainting. However, participants were not confined to these three features. We standardized the initial settings of Stable Diffusion to the “anything” model, “DPM++ 2M Karras” sampler, “Automatic” VAE model, and a size of 768*512 for each generation process. 

Throughout the generation process, participants could write prompts in any way they preferred, whether using complete sentences or single words, and had the freedom to employ any feature of Stable Diffusion to create and adjust images until they deemed the images to meet their expectations. After participants wrote Chinese prompts, researchers translated them into English using ChatGPT and then input them into Stable Diffusion. We made sure that the translation process did not add new words or meanings; researchers instructed ChatGPT to translate the prompts directly, preserving the original Chinese meaning and phrasing. All researchers used the same command-line template to input the prompts into ChatGPT for translation during the experiment. Researchers did not provide any guidance for the content creation, offering only technical support. Each participant completed two counterbalanced image generation processes for both familiar and unfamiliar cultural heritage. 

After each generation process, participants were invited to participate in a semi-structured interview. During the interview, participants were prompted to introduce the themes of the generated images and share their comments. Additionally, they were asked to describe and discuss the process of their collaboration with generative AI in creating the narrative.

\subsubsection{Data Recording}
During the process, we obtained participants' consent to record videos of the Stable Diffusion interface and the online meeting for the entire workshop. Participants were also required to copy the image and its related information and paste it into an online shared file. This served both as a means for them to track their output images and prompts during the process and for us to collect this information for further analysis.

\subsection{Data Analysis}

We employed thematic analysis to identify potential principal themes from the interview data. In analyzing the participants' interview data, we reviewed related prompts, generated images, and real-life photos to verify and understand the details they mentioned during the interviews, including how they altered their prompts, evaluated the generated images, and compared the images to actual cultural heritages. Three researchers conducted a round of independent open coding ~\cite{stol2016grounded} and discussed the resulting themes. The primary codes in the first round of coding encompassed AI's presentation of specific cultural features, new elements introduced by AI in narratives, participants' emotions in the narratives, describing specific scenarios in narratives, adapting narrative strategies based on AI, etc. We then went back to the data to conduct two additional rounds of focused coding ~\cite{stol2016grounded}, ultimately achieving agreement on the finalized themes. All quotes used in the paper were translated into English by the lead author and checked by the co-authors.

Furthermore, we organized the images generated by participants into sequences based on familiar and unfamiliar cultural heritages to explore differences in the quantity of images generated, the extent to which participants altered image elements, and other aspects.

\subsection{Participants}
A total of 20 individuals (Table 1) were recruited from various regions across China through online social platforms, with an age range of 18 to 35 years (M = 26.5, SD = 4.85). The participant group comprised 15 females and 5 males. Participants needed to create images of a familiar cultural heritage and another of an unfamiliar one. Therefore, our recruitment criteria required applicants to be very familiar with a cultural heritage site, such as a place in their hometown or a place where they had spent considerable time, and to have a good understanding of its appearance, culture, and other relevant aspects. Additionally, from the list of unfamiliar cultural heritage sites we provided, they needed to select at least one place they had never visited or knew very little about for creating images. We excluded participants who did not satisfy these criteria. Out of the 20 participants, only 2 had frequently used generative AI tools before; the rest had either never used them or had only used them a few times. Thus, most participants had limited experience using generative AI tools before the experiment. Regarding familiar cultural heritages, 13 out of 20 participants are local residents familiar with their cultural heritage, which they frequently visited. 1 participant lived in the city where their cultural heritage is located for several years. The remaining 6 participants have visited the familiar cultural heritage sites several times and possess considerable knowledge about them. Regarding unfamiliar cultural heritages, 19 out of 20 participants had never been to the unfamiliar cultural heritage sites we provided before the experiment; only P10 had visited the Star Ferry once, but she mentioned that she had merely passed by and had not entered, so she was not familiar with the place. They relied solely on pictures and words to learn about the unfamiliar places. Participants were compensated with a Stable Diffusion handbook containing feature demonstrations. Prior to participation, informed consent was obtained from all individuals. Additionally, all collected data was anonymized with the participants' permission to ensure their privacy.

\raggedbottom
\begin{table*}[htbp]
\centering
\caption {\label{tab:table1} Demographic Information of Participants}
\resizebox{\linewidth}{!}{
\begin{tabular} {lccccccccc}
\toprule
\textbf{ID}  &\textbf{Gender}   &\textbf{Age}  &\textbf{Education Background} &\textbf{\makecell{Familiarity with \\ generative AI tools}}   &\textbf{Region of Residence} &\textbf{Familiar Cultural Heritage} & \textbf{Familiarity} & \textbf{ \makecell{Unfamiliar \\Cultural Heritage}} & \textbf{Familiarity} \\
\midrule
$P1$  & Female & 18-25    & Humanities and Social Sciences & Used a few times & Jiangsu Province  & Ming City Wall & Local resident & Fruit Market & Never visited\\
$P2$  & Male    & 18-25   & Sciences and Engineering & Never used & Guangdong Province  & Da Fen Oil Painting Village & Local resident & Star Ferry & Never visited\\
$P3$  & Female    & 18-25    & Sciences and Engineering & Never used & Guangdong Province & \makecell{Monument to the Autumn \\ Harvest Uprising} & Local resident & Fruit Market & Never visited\\
$P4$  & Female    & 26-35  & Sciences and Engineering & Never used & Guangdong Province & Dunhuang & Visited several times & Fruit Market & Never visited \\
$P5$  & Male    & 26-35    & Humanities and Social Sciences & Frequently uses & Beijing &The Forbidden City & Local resident & Mid-level Central & Never visited \\
$P6$  & Female  & 26-35   & Business  & Used a few times & Guangdong Province & Nantou Ancient Town & Local resident & Fruit Market & Never visited \\
$P7$  & Female  & 18-25  & Humanities and Social Sciences & Used a few times & Shandong Province & Haiyun Temple Tangqiu Festival & Local resident & Kowloon Walled City & Never visited \\
$P8$  & Male    & 18-25  & Sciences and Engineering & Never used & Guangdong Province & Shantou Small Park & Local resident & Lui Seng Chun & Never visited \\
$P9$  & Female  & 18-25  & Humanities and Social Sciences & Used a few times & Zhejiang Province &Grand Buddha Temple & \makecell{lived for \\a few years} & State Theatre & Never visited \\
$P10$  & Female  & 18-25   & Humanities and Social Sciences & Used a few times &Guangdong Province &Chang'an Road & Local resident & Star Ferry & Visited once \\
$P11$  & Female  & 18-25   & Humanities and Social Sciences & Used a few times & United Kingdom &York Minster & Visited several times & State Theatre & Never visited \\
$P12$  & Female  & 18-25  & Humanities and Social Sciences & Used a few times & Guangdong Province &Chaozhou Shadow Play & Visited several times & Ding Ding & Never visited \\
$P13$  & Male  & 18-25  & Humanities and Social Sciences & Frequently uses &Guangdong Province &Guanyin Mountain & Local resident & Kowloon Walled City & Never visited \\
$P14$  & Female    & 18-25  & Sciences and Engineering & Used a few times & Guangdong Province &Nanjing Massacre Memorial Hall & Visited several times & Central Market & Never visited \\
$P15$  & Female  & 26-35   & Humanities and Social Sciences & Never used & Jiangsu Province &Bell Tower & Local resident & Lui Seng Chun & Never visited\\
$P16$  & Male  & 26-35  & Humanities and Social Sciences & Never used & Shanghai &Suzhou Gardens & Visited several times & Ding Ding & Never visited \\
$P17$  & Female  & 18-25  & Humanities and Social Sciences & Used a few times  & Shanxi Province & Shanxi University Sculptures & Local resident & Yau Ma Tei & Never visited \\
$P18$  & Female  & 18-25   & Media and Communication  & Never used &Shandong Province &Laoshan Taoist Temple & Visited several times & Yick Cheong Building & Never visited \\
$P19$  & Female  & 18-25   & Humanities and Social Sciences & Never used &Henan Province &Earthen Wall Grass House & Local resident & Yick Cheong Building & Never visited \\
$P20$  & Female  & 26-35  & Humanities and Social Sciences & Never used & Guangdong Province &Yuyin Shanfang under Canton Tower & Local resident & Yau Ma Tei & Never visited\\
\\
\bottomrule
\end{tabular}
}
\label{table:01}
\Description{Demographic Information of Participants}
\end{table*}

\section{Results}\label{sec:Results}

\subsection{Strategies Participants Adopted When Creating a Narrative Related to Familiar Cultural Heritage in Collaboration with Generative AI}
The results of thematic analysis indicated that, when collaborating with Stable Diffusion to generate images of familiar cultural heritage, participants adopted similar creation strategies. These findings present the strategies participants tended to adopt when expressing their personal narratives of familiar cultural heritage through AI collaboration. 

\subsubsection{Three creation strategies for generating images of familiar cultural heritages}
Based on participants' descriptions of their creative themes in the interviews, we found that their approaches varied: some participants tended to objectively recreate the actual state and characteristics of cultural heritages, some added their own memories, experiences, and emotional connections to depict a scenario within the cultural heritage, and others used a creative approach to make interesting creations based on the cultural heritage. We categorized these three creative approaches as (1) depicting realistic scenes, (2) recreating memorable scenarios, and (3) expressing exploratory creation. We then placed them on a spectrum from objectively depicting heritage sites to subjectively expressing interpretations to indicate whether participants' creative attitudes were more objective or subjective (see fig.2).

\begin{figure}[htbp]
  \centering
  \includegraphics[width=0.9\textwidth]{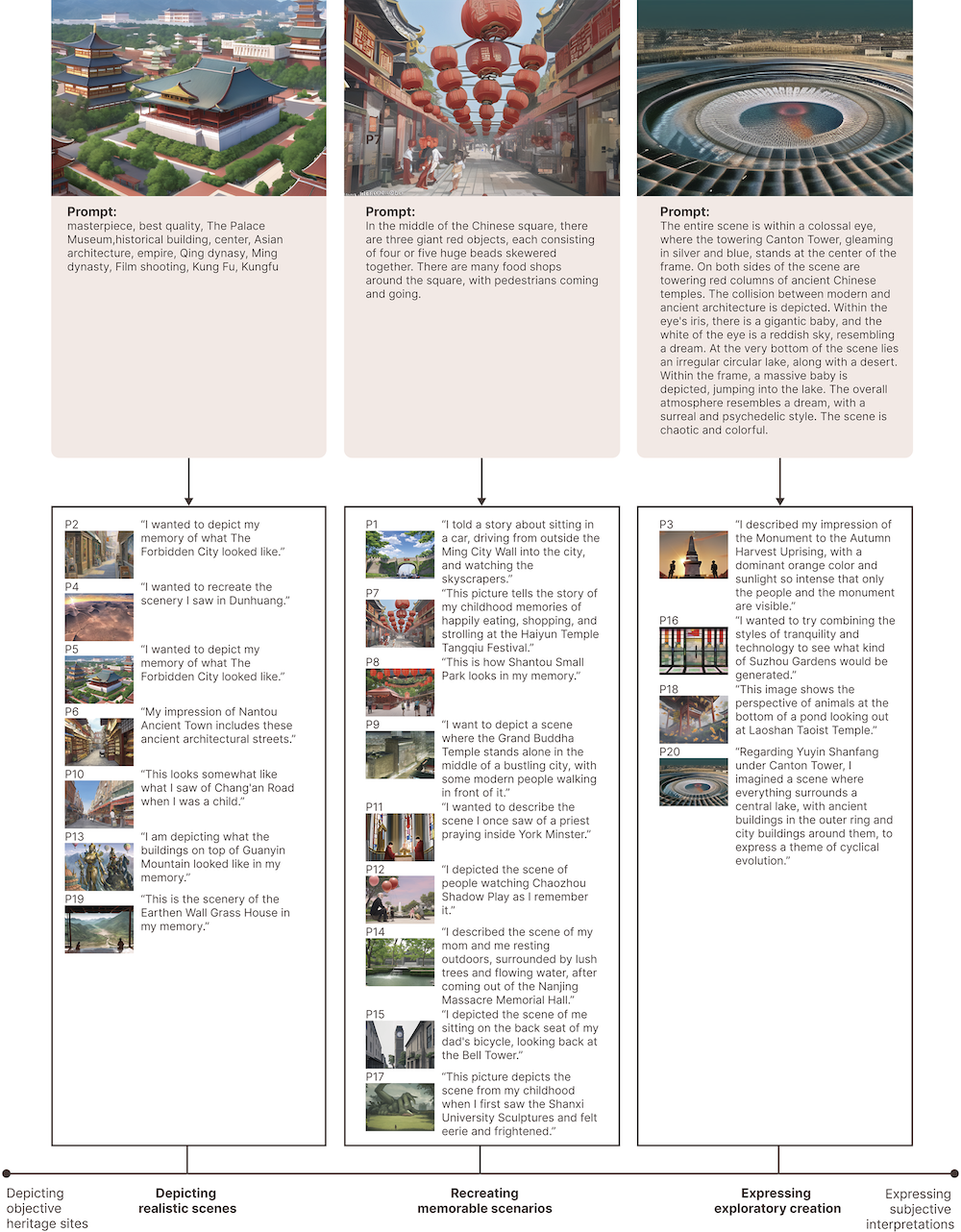}
  \caption{Three creation approaches used by participants, each corresponding to the images they ultimately generated. These three approaches encompass depicting realistic scenes, recreating memorable scenarios, and expressing exploratory creation.}
  \label{P1}
  \Description{Caption}
\end{figure}

Participants who aimed to depict realistic scenes typically described specific details of the scene, such as the visual angle (P4), the appearance of the architecture (P2, P5), and the environmental characteristics (P6, P10, P13, P19). They compared the details of the output images with the real scenes that they had observed, and corrected any inaccuracies to make the images more closely resemble what they remembered seeing.

Those participants intending to recreate memorable scenarios often infused their emotions and memories into the images, focusing on impressions (P1, P7, P9, P11, P12) or events (P8, P14, P15, P17) related to cultural heritages. In order to convey their feeling associated with the heritages, participants adjusted color schemes, sizes and shapes of components, composition, and more. The goal was not necessarily to replicate the real scene but to capture their impressions. For instance, P9 explained, "\textit{What I wanted to show was how the Dafo Temple looked to me when I first saw it. It's this old Buddhist building standing tall in the middle of a busy modern city. I wanted to capture how the old and new mix and clash.}". Participants often placed a small figure in the scene as a symbol of themselves to anchor their past selves in the narrative. As P7 noted, "\textit{I added a character in the scene, hoping to depict myself in that past scenario.}"

Participants expressing exploratory creation pursued beyond the cultural heritages, presenting fantastical and imaginative ideas primarily rooted in their personal creativity rather than reality or memory (P3, P16, P18, P20). For example, P20 told us: "\textit{What I initially envisioned was a cyclic pattern, where everything undergoes a continuous cycle of transformation.}" They embraced bold attempts, viewing the imaginative results generated by AI as creative inspiration, as expressed by P20: "\textit{I think one good thing about AI is that it can stimulate your imagination; it can come up with things that people wouldn't think of. So, as I proceed, I hope it can be more daring.}"

\subsubsection{Participants may change their creative strategies after gaining some understanding of the AI}
Some participants, after seeing AI-generated images, would have a general understanding of the possible creative results, thus changing their creative methods to align with their perception of the AI's characteristics (P3, P17, P20). For example, P20 changed her creation strategy from "depicting realistic scenes" to "expressing exploratory creation" during the process (see fig.3). She initially added a surreal element to a prompt that originally described an objective scene by including the phrase "\textit{inside a massive eye,}" and the resulting image transformed from a realistic style to a more dreamlike one. She then further refined the description of the massive eye and added more mystical terms such as "\textit{dream,}" "\textit{surreal,}" and "\textit{psychedelic.}" This gradually shaped the image into the desired composition, with a circular area resembling an eye at the center, surrounded by buildings. She told us, 
\begin{displayquote}
"\textit{At first, I just wanted to present the lake, ancient buildings, and the Canton Tower, but when I found it difficult to achieve, I changed my approach. I chose one element that could be extended and added some of my favorite elements. I think the good thing about AI is that it can stimulate your imagination, coming up with things that people might not think of. So, the further I went, the more I hoped it could be bolder. As a result, the later works almost resemble the universe, not just limited to my imagination.}"
\end{displayquote}

\begin{figure}[htbp]
  \centering
  \includegraphics[width=0.7\textwidth]{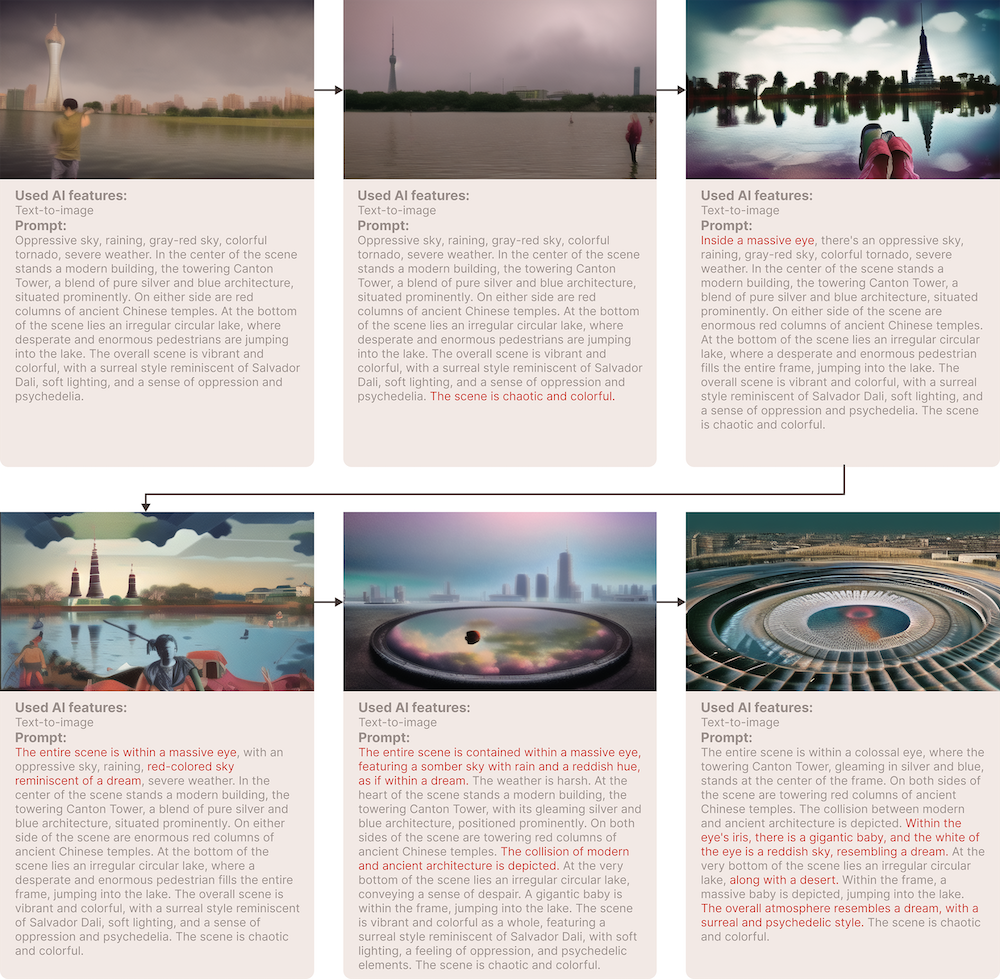}
  \caption{P20's generated images and prompts. She gradually incorporated many surreal elements into her prompts.}
  \label{fig3}
  \Description{Caption}
\end{figure}

\subsection{Generative AI's Strengths in Assisting Personal Narrative Creation About Familiar Cultural Heritage}
Through thematic analysis, we identified three types of assistance that AI provides to personal narratives: illuminating personal narratives, amplifying personal narratives, and reinterpreting personal narratives.

\subsubsection{Illuminating Personal Narratives}
AI can assist participants in constructing their personal narratives by iteratively generating images and gradually awakening their memories. Since our participants are experts in the cultural heritage they are familiar with, they were able to evaluate and correct inaccuracies in the generated images. However, they were not expert in storytelling, thus they initially did not have a clear anticipation of the visual outcome to be generated, only some scattered ideas (P1, P9, P11, P14). However, the rapid presentation of AI-generated results enabled participants to continuously experiment with new keyword combinations until achieving their desired effect. For example, P11 initially focused on describing a building but gradually incorporated diverse narrative elements\add{,} such as the Bible and priests\add{,} to enrich the storytelling aspects of the image (see Fig. 4). As she added more detailed descriptions in the prompts, such as "\textit{priest in a red robe}" and "\textit{a Bible on the table,}" the images evolved from merely showcasing the church building to depicting a narrative scene with a priest praying. Although the image did not perfectly capture details from the prompt, such as "\textit{three worshippers,}" it still effectively enhanced the overall storytelling aspect that P11 intended. Another example is P1, who used the inpainting feature to continuously adjust the traffic volumes on the road (see Fig. 5). She initially used a brush to erase the road area and then input "\textit{cars,}" but no cars appeared in the image. She then tried "\textit{cars on the road}" to depict cars driving on the road, which successfully generated cars. Afterwards, she added "\textit{bus}" to the prompt, but the resulting image had too much traffic, making it feel crowded. Ultimately, she chose the second generated result. This process illustrates how stable diffusion quickly provided P1 with various results, helping her find the desired visual effect. 

\begin{figure}[htbp]
  \centering
  \includegraphics[width=1\textwidth]{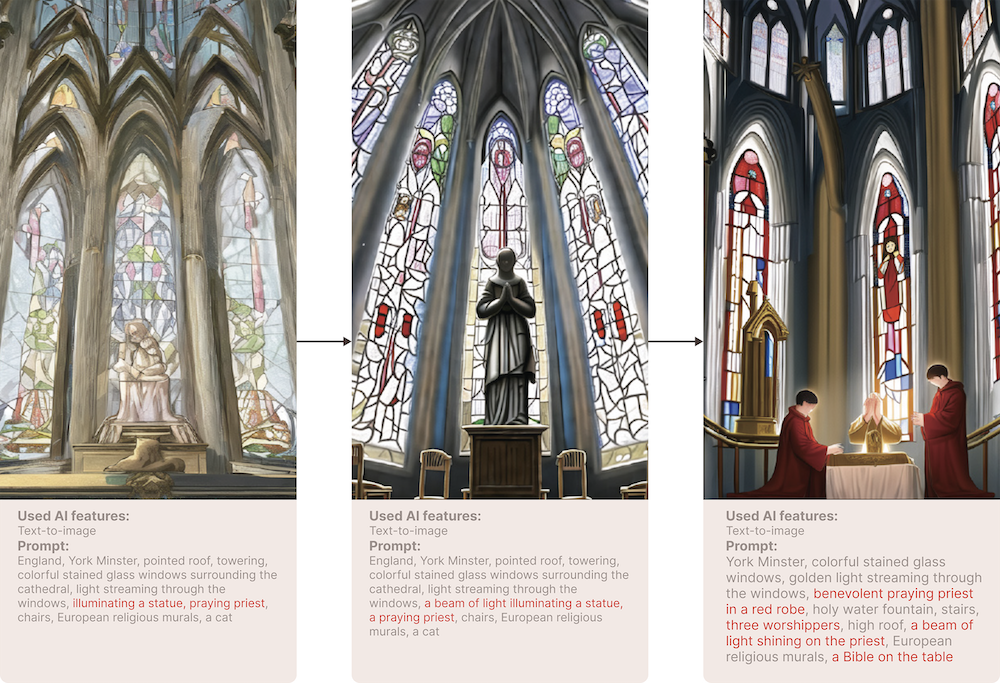}
  \caption{P11's generated images and prompts evolved from her initial goal of depicting a building to gradually incorporating various narrative elements.}
  \label{fig4}
  \Description{Caption}
\end{figure}

\begin{figure}[htbp]
  \centering
  \includegraphics[width=1\textwidth]{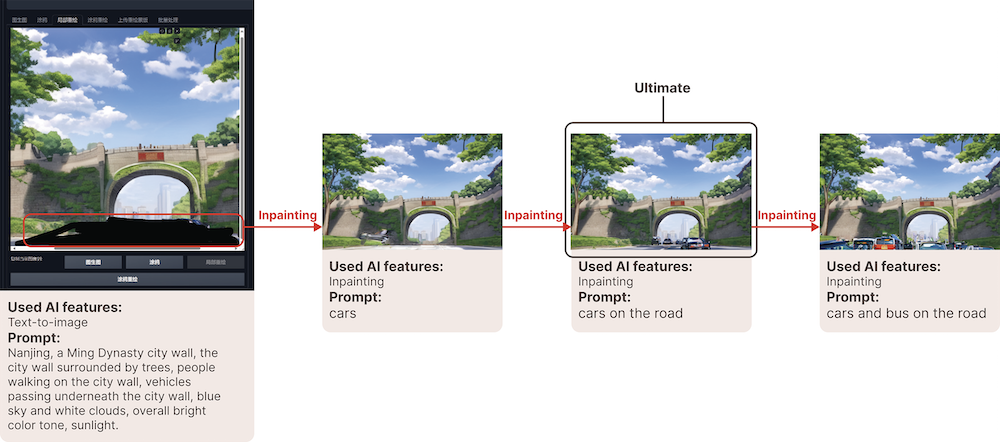}
  \caption{ P1's inpainting interface and the generated images. She tested a number of scenarios with different traffic volumes on the road and finally choose one from them.}
  \label{fig5}
  \Description{Caption}
\end{figure}

Some participants' memory of the cultural heritage is vague, making it challenging for them to recall specific details when generating images initially without any references. In the process, The AI-generated images concretely portrayed the blurred images in participants' minds, enabling them to compare these visuals with their own memories. Interestingly, some of the AI's randomly generated unexpected elements occasionally prompt them to recall additional details and emotions associated with the cultural heritage (P1, P14, P17, P19). For example, P17 told us that the generated image truly evoked her childhood fear of that sculpture (see fig.6):
\begin{displayquote}
"\textit{At first, I said it was 'really, really scary' when I saw it as a child, but now, it's just three words for me. However, when I was a child, those three words felt like the world was about to end. During the generation process, for instance, I added the keyword 'thrilling.' As a result, the imagery became more abstract and distorted. Many bizarre and grotesque figures emerged in the picture, and I really felt that those three words, 'really, really scary,' came alive.}"
\end{displayquote}
\begin{figure}[htbp]
  \centering
  \includegraphics[width=1\textwidth]{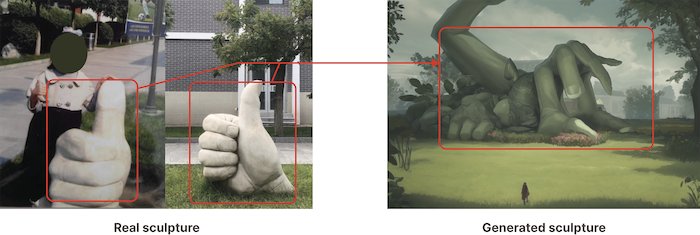}
  \caption{The photograph of the authentic sculpture (provided by P17) is compared with the image generated by P17.}
  \label{P4}
  \Description{Caption}
\end{figure}

\subsubsection{Amplifying Personal Narratives}
AI amplify personal narratives by introducing supplementary narrative elements or suggesting alternative visual representation, which, after a selection process, eventually become integrated into the final image. The generated images incorporated additional elements not initially specified by the participants in their keywords, occasionally unexpectedly enhancing the original narrative (P6, P10, P18). For example, in the image of the Qilou (a traditional arcade buildings found in Southern China) generated by P10, AI introduced numerous shops, products, and customers. While this depiction differed from her existing impression of the Qilou scene, it significantly enriched her narrative (see fig.7). P10 told us:
\begin{displayquote}
"\textit{The generated images are somewhat different from my memories; they appear more vivid, and the entire scene seems crowded. In reality, the pathway wasn't as crowded. It gives the impression that there are many more shops and their products inside, making me feel happy and content walking through it. Because it used to be lively here, and now it feels deserted.}"
\end{displayquote}

\begin{figure}[htbp]
  \centering
  \includegraphics[width=1\textwidth]{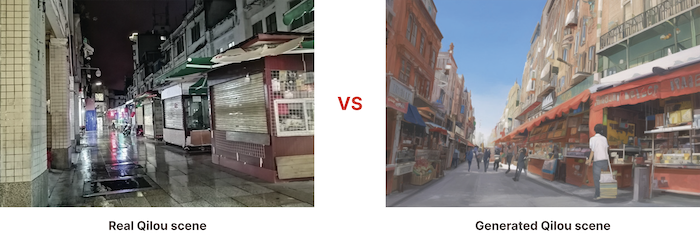}
  \caption{The photograph of the authentic Qilou scene (sourced from the internet) and the image of Qilou generated by P10.}
  \label{P5}
  \Description{Caption}
\end{figure}

Additionally, AI might present an image with visual representations different from what the participant initially intended, but it more effectively captures their main idea (P1, P13, P15, P16, P17). For instance, P15 experimented with various descriptions in the prompt, such as "\textit{a crowd of people riding bicycles}" and "\textit{a father riding a bicycle carrying a little girl.}" Most of the images were depicted from a third-person perspective. However, one randomly generated image, which appeared to be from a first-person perspective, reminded her of how she used to see the clock tower while sitting on the back of her father's bicycle. She felt that this unexpected image better expressed her idea (see Fig. 8):
\begin{displayquote}
"\textit{I've been struggling with depicting a scene of someone riding a bicycle. Initially, I wanted to show it from a third-person perspective. Later, I found that the current generated perspective actually resembles what I intended. It's as if I'm still sitting on the back seat of that bicycle, looking back at the scene of the building.}"
\end{displayquote}

\begin{figure}[htbp]
  \centering
  \includegraphics[width=1\textwidth]{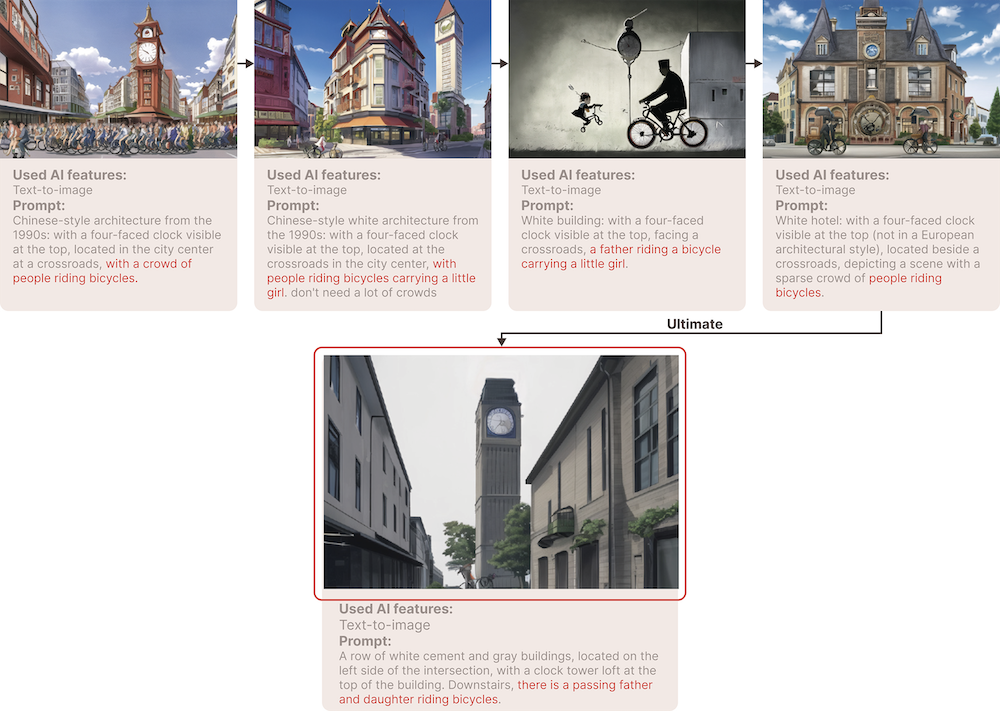}
  \caption{The images of the clock tower scene generated by P15, she chose one that happened to be generated, resembling a first-person perspective.}
  \label{P6}
  \Description{Caption}
\end{figure}

\subsubsection{Reinterpreting Personal Narratives}
Some of the unconventional outcomes produced by AI appear thought-provoking to certain participants, prompting them to reconsider their narratives and, in turn, come up with new interpretation (P13, P18, P20). For instance, when AI misunderstood "\textit{red moon}" and generated a "\textit{red eyeball}" (see fig.9) it sparked new thoughts for P18:
\begin{displayquote}
"\textit{Usually, when we look at this painting, we perceive the scene from our own perspective. However, with an eyeball appearing on it, it feels like the scene is staring back at us, creating a shift in perspective.}"
\end{displayquote}

\begin{figure}[htbp]
  \centering
  \includegraphics[width=0.6\textwidth]{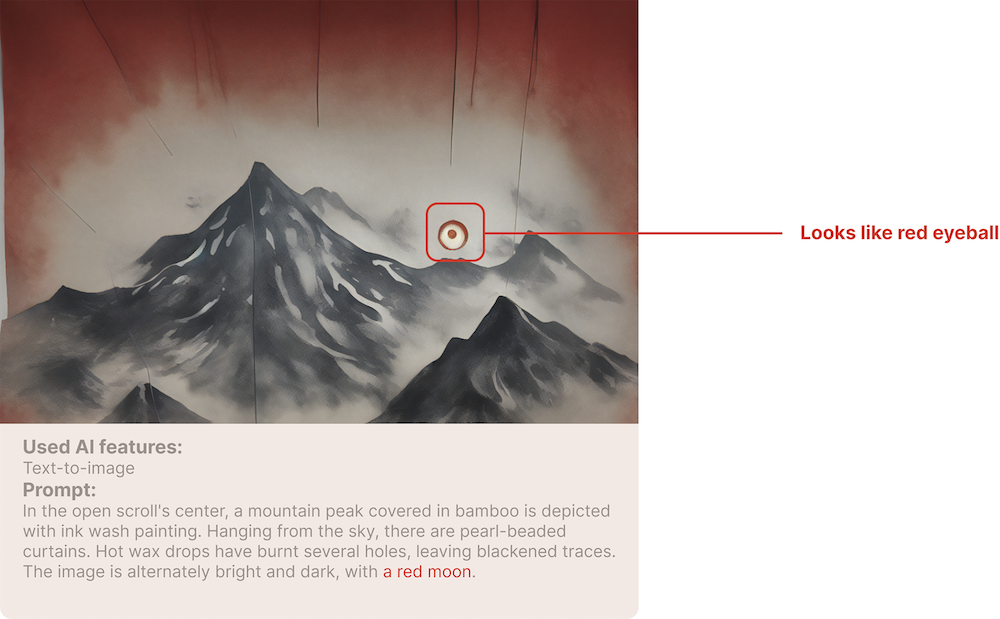}
  \caption{In a image generated by P18, the red moon appears to resemble a red eyeball.}
  \label{P7}
  \Description{Caption}
\end{figure}

The image of the deity generated by AI is very different from the one P13 had in mind (see fig.10), but he finds it interesting, he told us:
\begin{displayquote}
"\textit{What's kind of cool is how the AI shows the deity in different ways. Normally, we all have a fixed idea of what gods look like. They have a set appearance in our minds. But the AI comes up with different images, like how their face or clothes look. This made me wonder, why do we always picture deities in a specific way?}"
\end{displayquote}

\begin{figure}[htbp]
  \centering
  \includegraphics[width=1\textwidth]{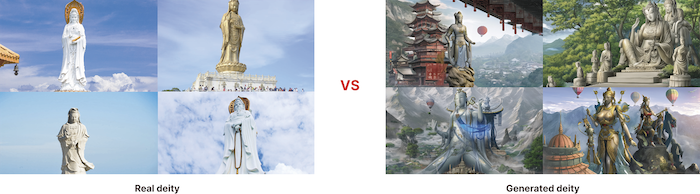}
  \caption{The photograph of the deity (sourced from the internet) and the image of deity generated by P13.}
  \label{P8}
  \Description{Caption}
\end{figure}

Overall, these findings indicated that AI can assist participants to varying degrees in completing their personal narratives about cultural heritage. It can help them clarify their narrative thoughts, enabling them to recall more details related to cultural heritage. AI can also suggest supplementary narrative elements and present ideas to help participants more comprehensively express their themes. Further, the uncertainty introduced by AI may even alter participants' original understanding, leading to the creation of new and more meaningful narratives.

\subsection{Generative AI's Limitations in Personal Narrative Creation About Familiar Cultural Heritage}

Generative AI presents limitations in familiar cultural heritage narrative creation, including limitations in meeting detailed participant requirements, struggles in depicting specific cultural features, and cultural bias in favor of Western styles. 

\subsubsection{Generative AI's Limitation in Meeting Detailed Requirements for Familiar Cultural Heritage Narrative}

The images generated by AI at times fall short of meeting participants' detailed requirements (P3, P15). Consequently, certain elements in the scenes may lack relevance to both the cultural heritage and the narrative intended by the participants, potentially resulting in an incomplete representation of their storytelling and leading to misunderstandings. For instance, when P3 attempted to describe a scene with five soldiers standing beside the monument, the AI consistently portrayed the soldiers standing on top of the monument (see fig.11). As P3 said:
\begin{displayquote}
"\textit{Even though I wrote that all five people are standing beside it, AI always insists on placing them on top of the monument, but originally, there wouldn't be anyone standing on it. It merges the people with the monument, thinking that I meant people are also part of the monument.}"
\end{displayquote}

\begin{figure}[htbp]
  \centering
  \includegraphics[width=1\textwidth]{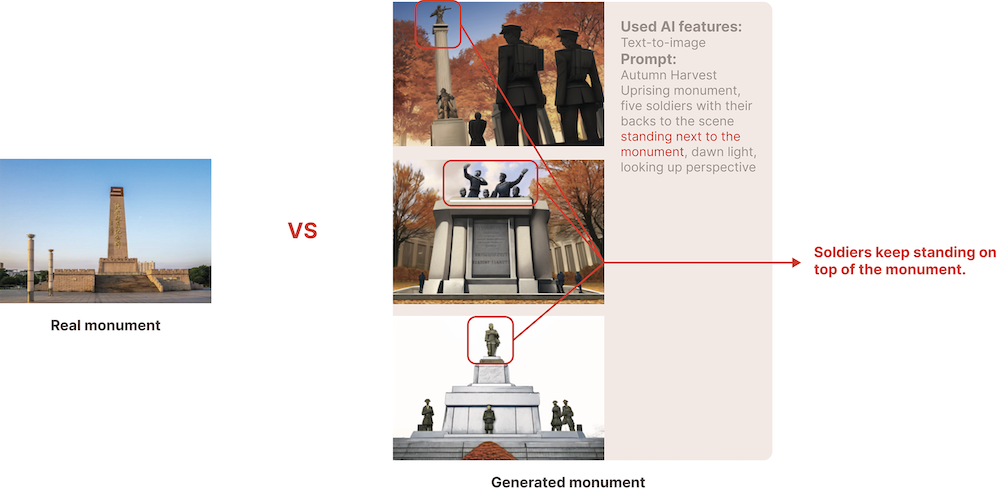}
  \caption{The photograph of the monument (sourced from the internet) and P3's AI-generated image and prompt for the same monument.}
  \label{P14}
  \Description{Caption}
\end{figure}

However, some participants became aware of this limitations of AI during the generation process, prompting them to modify their expectations and creative strategies (P5, P7, P11, P17, P20). For instance, P7 mentioned,  "\textit{After the first round of generating, I pretty much got the vibe, so I moved on to the next step without making too many adjustment. Also, when I'm adjusting, I don't set my expectations too high.}" Recognizing both the weaknesses and strengths of AI, participants discovered that adjusting their generation strategies sometimes yielded unexpectedly positive outcomes. For instance, P17, after gaining a better understanding of AI's strengths, began incorporating abstract descriptions into her prompts, such as "\textit{an overall atmosphere of suspense,}" to make the images more surreal and uncanny (see Fig.12). P17 told us:
\begin{displayquote}
"\textit{At first, I'd get pretty frustrated. I mean, the inputs I gave were so spot-on, and then it gives me these completely different images. Later on, I figured since it can't nail down a specific image, why not take a more abstract approach myself? I tried describing things in a different way, turning the real-world look into more of an imagined or abstract feel. As we gradually got in sync, I started feeling like it's a kind of relationship adjustment between me and the machine. Now, I'm quite satisfied with the current image. If we keep refining, there might even be a chance to generate an even better result.}"
\end{displayquote}

\begin{figure}[htbp]
  \centering
  \includegraphics[width=0.9\textwidth]{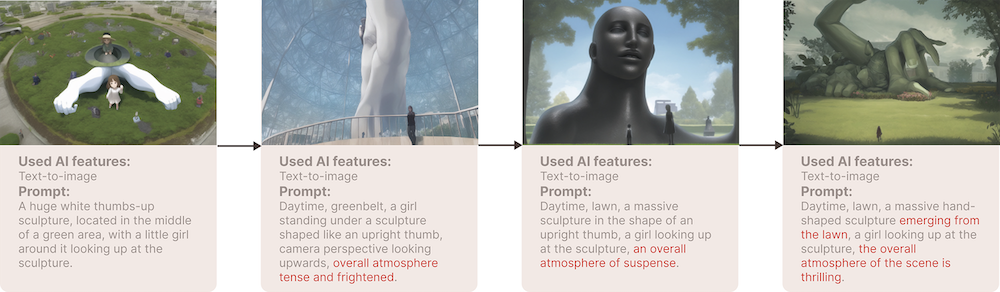}
  \caption{The AI-generated images by P17 and the accompanying input prompts. She gradually added more abstract descriptions within the prompts.}
  \label{P15}
  \Description{Caption}
\end{figure}

\subsubsection{Generative AI's Limitation in Presenting Specific Cultural Features}
Cultural heritages encompass unique cultural forms, and participants were particularly keen on accentuating those distinctively culturally significant features in the images. However, AI faced challenges in comprehending or portraying those specific cultural images due to a lack of relevant datasets. This inadequacy led participants to feel that the images lacked crucial cultural elements (P7, P13, P16, P17, P20). For instance, P7 attempted to use various words to describe the image of "Tangqiu" (a traditional Chinese food of sugar balls). She tried terms like "\textit{sugar ball,}" "\textit{gobstopper,}" and "\textit{jawbreaker,}" but AI consistently failed to depict the distinctive appearance of Tangqiu as envisioned by her (see fig.13). P7 told us:
\begin{displayquote}
"\textit{The most significant issue is with the key term 'Tangqiu'. When describing the Tangqiu gathering, it's essential to highlight the main aspect of 'Tangqiu'. However, by the end of the generation process, the image of 'Tangqiu' still did not come across as well as I had hoped in my mind.}"
\end{displayquote}

\begin{figure}[htbp]
  \centering
  \includegraphics[width=1\textwidth]{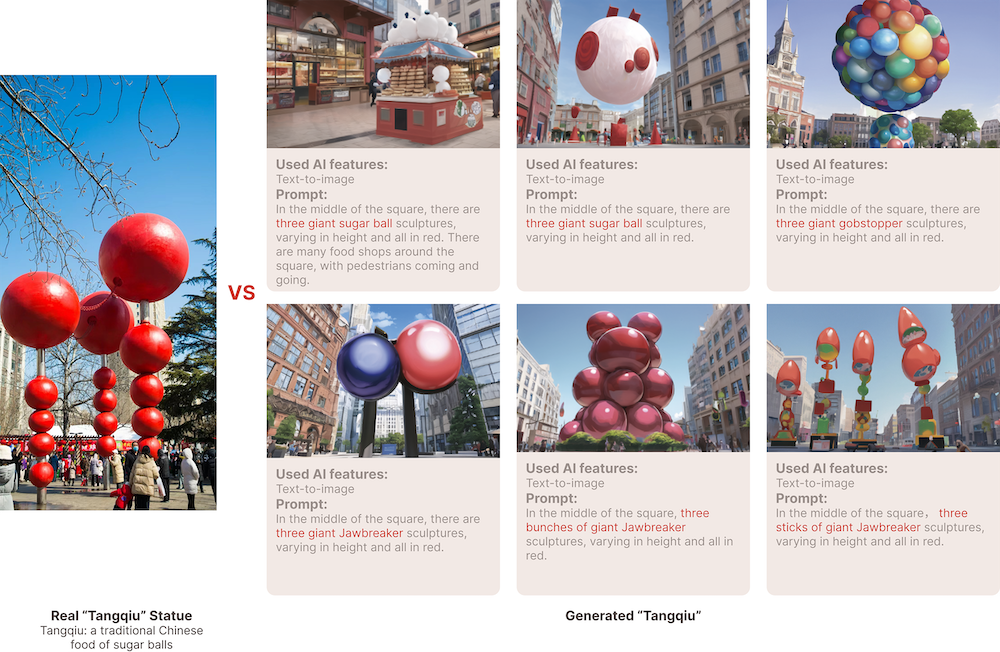}
  \caption{The photograph of "Tangqiu" (sourced from the internet), along with the AI-generated images and prompts created by P7 for "Tangqiu".}
  \label{P9}
  \Description{Caption}
\end{figure}

\subsubsection{Generative AI's Cultural Bias}
We have observed that AI tends to generate buildings with a predominantly Western style, often diverging significantly from the traditional architectural aesthetics found in China. For example, P15 noted that the buildings in the generated images mostly lean towards a European style, lacking the sense of historical periods and local characteristics inherent in Chinese architecture (see fig.14). She commented, 
\begin{displayquote}
"\textit{The pictures we generated along the way mostly look like European-style buildings, which is totally different from what I remember. I recall there being modern buildings with some era-specific and local characteristics. Although these buildings in my hometown doesn't have that ancient history spanning dynasties, it has witnessed changes over the past thirty years or so.}"
\end{displayquote}

\begin{figure}[htbp]
  \centering
  \includegraphics[width=0.8\textwidth]{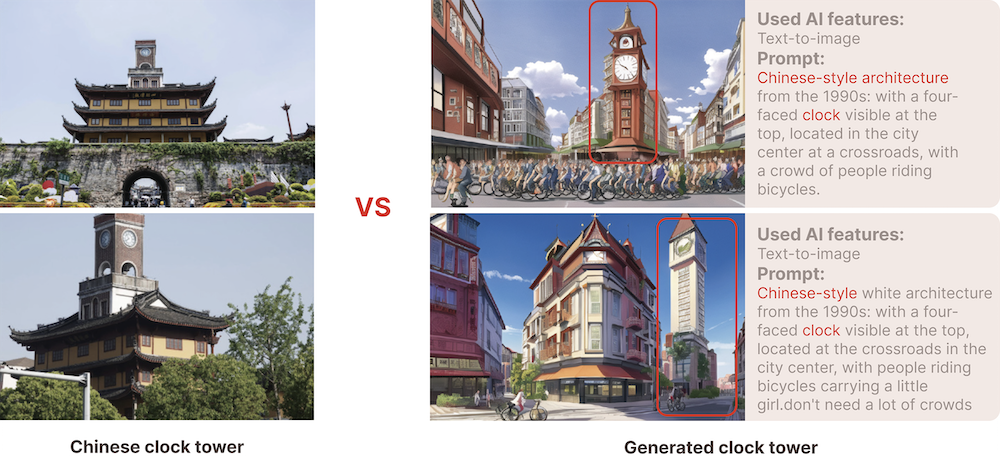}
  \caption{The photograph of the Chinese clock tower (sourced from the internet) and the AI-generated clock tower by P15.}
  \label{P9}
  \Description{Caption}
\end{figure}

In addition, AI often relies on stereotypical depictions of Chinese buildings and characters. For instance, when P7 described a unique and charming Chinese temple street for the "Tangqiu" festival, the resulting architecture in image (see fig.15) gave her the impression that "\textit{The buildings it's creating lean more towards a flashy, colorful vibe, not quite that serene charm you'd find in a temple. It's more like something you'd see in a Chinatown.}" Moreover, extending beyond architecture, when P7 tried to use the inpainting feature to generate an image of a young girl placed on the street, the AI defaulted to creating a girl with a Western appearance when she input "\textit{a little girl.}" Even after modifying the prompt to "\textit{Chinese little girl,}" the generated characters still retained Western features (see Fig. 16).

\begin{figure}[htbp]
  \centering
  \includegraphics[width=1\textwidth]{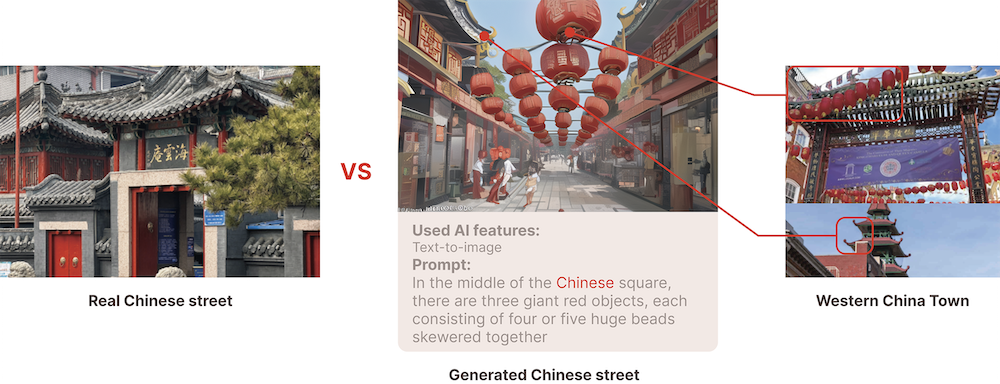}
  \caption{The photograph of the real Chinese street (sourced from the internet) and P7's AI-generated image and prompt for the Chinese street. The AI-generated image specifically captures the style of a Western Chinatown.}    
  \label{P12}
  \Description{Caption}
\end{figure}

\begin{figure}[htbp]
  \centering
  \includegraphics[width=0.6\textwidth]{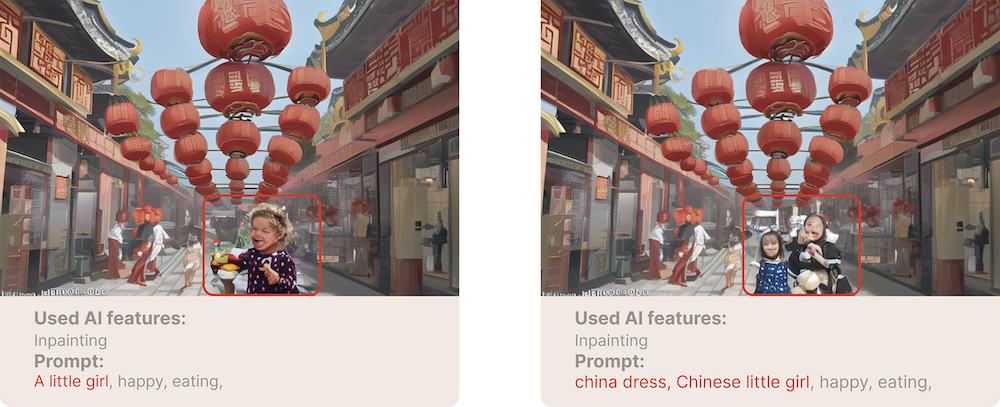}
  \caption{In the images generated by P7, the young girls display characteristics typical of Western children.}
  \label{P13}
  \Description{Caption}
\end{figure}


\subsection{The Impact of Lacking Local Knowledge on Using Generative AI to Create Images of Unfamiliar Cultural Heritages}

To understand how people's expertise in local culture heritage influences their generative process and outcomes when collaborating with AI in storytelling, we had participants generate additional images based on provided information about a cultural heritage they were not familiar with. Our findings indicated that, owing to a lack of professional knowledge about the specific cultural heritage, participants were more inclined to accept the generated results, in contrast to their hesitancy in generating content related to familiar cultural heritage. Also, the generated images were more likely to exhibit bias, and participants' perception of the cultural heritage is more easily influenced by the generated images.

\subsubsection{The Sequence of AI-Generated Images About Familiar and Unfamiliar Cultural Heritage}

\begin{figure}[htbp]
  \centering
  \includegraphics[width=1\textwidth]{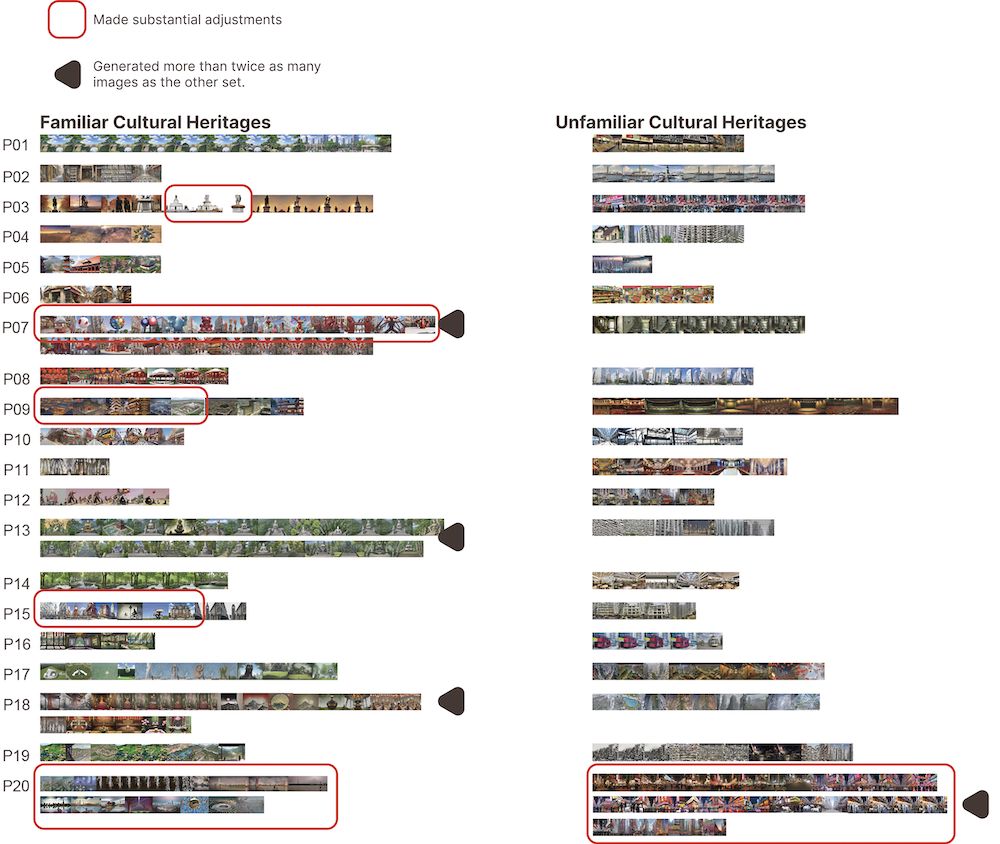}
  \caption{Sequence of output images generated by 20 participants}
  \label{P13}
  \Description{Caption}
\end{figure}

Figure 17 illustrates the sequence of output images generated by 20 participants, comparing the image sequences of familiar and unfamiliar cultural heritages. The red frames highlight images that were significantly altered by participants in terms of color, structure, and elements. The black arrows indicate the image sequences that contain more than twice as many images as the other set (familiar/unfamiliar cultural heritages). Within the context of familiar cultural heritages, participants consistently adjusted output images. Notably, P3, P7, P9, P15, made substantial adjustments to the generation process of familiar cultural heritages. However, only P20 made a substantial adjustments in the generation of unfamiliar cultural heritages, as evidenced by the red frame in Figure 17 . Furthermore, participants P7, P13, and P18 produced twice as many images of familiar cultural heritages compared to unfamiliar ones, as indicated by the black arrows in Figure 17 . This behavior can be attributed to their in-depth familiarity with these heritages, leading them to continuously modify details of the output images to align with their memories. In contrast, when generating images of unfamiliar cultural heritages, participants provided general descriptions and made fewer modifications to the output images. Due to a lack of knowledge about these cultural heritages, participants exhibited a higher level of acceptance towards the details generated by AI.

\subsubsection{Participants' Creative Strategies for Unfamiliar Cultural Heritage.}
When describing unfamiliar cultural heritage, participants depicted the architectural features they see in the available materials (see fig.18). For example, P1 aimed to convey the appearance of "\textit{a fruit market located among skyscrapers}" based on the information seen in the images. She mentioned, \textit{“I think it matches the characteristics of this fruit market. Although subjectively, I don't like this composition, objectively, I think it expresses what it wants to express. It is a run-down wholesale market within the city of Hong Kong”} Additionally, participants often incorporate impressions of the place formed through other channels. For instance, P5 integrated the "\textit{cyberpunk}" impression of Hong Kong formed from movies into his creation. He mentioned, "\textit{The second picture resembles many of those scenes in sci-fi movies,. It feels like, even though it's crowded, it could be an extreme form of a cyberpunk future.}"

\begin{figure}[htbp]
  \centering
  \includegraphics[width=1\textwidth]{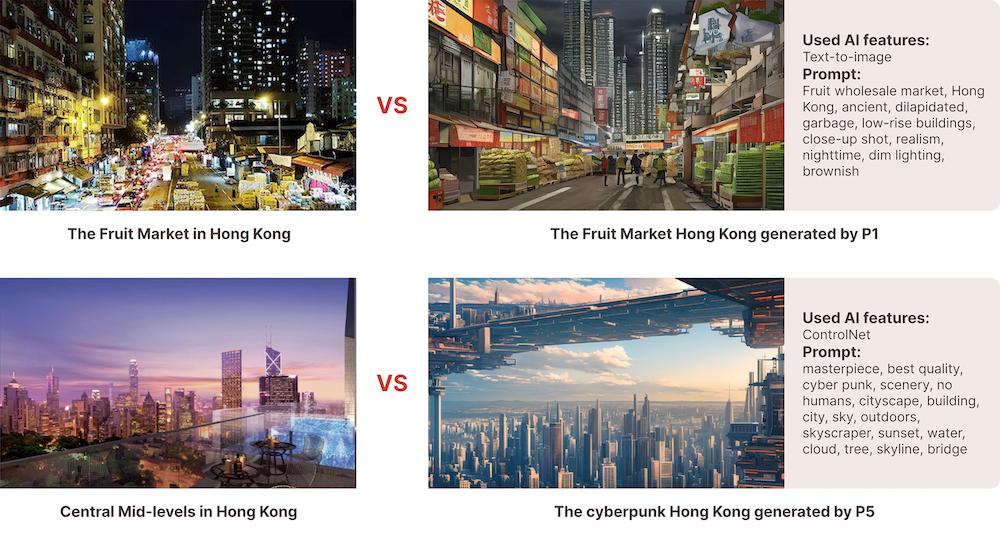}
  \caption{
The photograph of the Fruit Market and the Central Mid-levels in Hong Kong (sourced from the internet) and P1 and P5's AI-generated image and prompt for the same place.}
  \label{P1+P5}
  \Description{Caption}
\end{figure}

\subsubsection{Biases in Participants' AI-Generated Images of Unfamiliar Cultural Heritage}
When generating images related to unfamiliar cultural heritage, participants frequently incorporated their generalized impressions of a place (P5, P6, and P12), which may deviate from reality. As a result, participants might inadvertently introduced their own stereotypes into the images. For instance, when P5 described Central Mid-levels in Hong Kong, a place he had never visited, he infused it with "\textit{cyberpunk}" features and believed that the generated image aligned with his perception of Hong Kong as "\textit{crowded}" (see fig.19). He mentioned:
\begin{displayquote}
"\textit{I've been to Hong Kong a few times, and the vibe there is quite stifling, humid, and crowded. The first generated image, even though it has this cyberpunk and sci-fi feel, still captures that crowded essence. I use the term 'cyberpunk' because I've seen a lot of sci-fi movies and shows that are set in Hong Kong and Tokyo.}"
\end{displayquote}
\begin{figure}[htbp]
  \centering
  \includegraphics[width=1\textwidth]{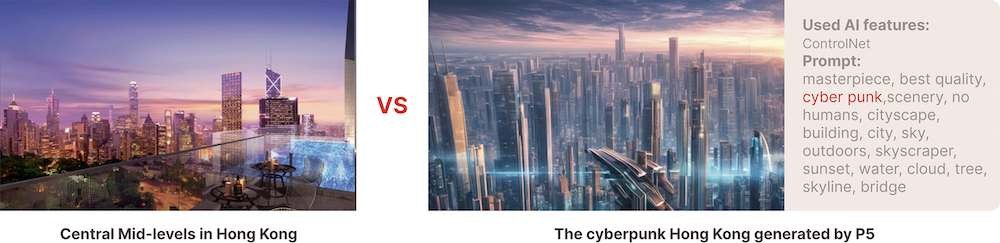}
  \caption{The photograph of the Central Mid-levels in Hong Kong (sourced from the internet and P5's AI-generated image and prompt for the same place. P5 incorporated "cyberpunk" elements into the image.}
  \label{P16}
  \Description{Caption}
\end{figure}

Additionally, participants' lack of familiarity with the location hinders their ability to discern potential cultural biases within the generated images, making it challenging for them to identify and modify such biases. For example, in the cultural heritage related to the fruit market site that we provided, the most typical architectural feature is the rooftop adorned with old-style three-dimensional calligraphy on stone signs. However, during her creative process, P1 failed to notice that the buildings in the image lacked this distinctive cultural feature, appearing too commonplace (see fig.20). Furthermore, her perception of the place upon seeing the image was "\textit{I think it's kind of like an urban village where young people have probably gone out to work, and there might be some elderly folks here, setting up stalls to sell fruits.}", while, in reality, this fruit market serves as a local central hub for the import of fruits from around the world. Many vendors engage in trading and transportation here, and it is not a backward urban village.
\begin{figure}[htbp]
  \centering
  \includegraphics[width=1\textwidth]{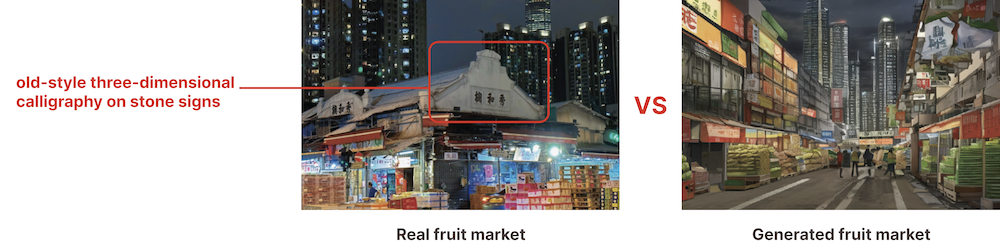}
  \caption{The photograph of the fruit market (sourced from the internet) and P1's AI-generated image for the same place. The most typical architectural feature of the fruit market is the rooftop adorned with old-style three-dimensional calligraphy on stone signs.}
  \label{P17}
  \Description{Caption}
\end{figure}

Due to their limited knowledge of the location, the generated images have the potential to influence and shape their impressions of the place, possibly introducing cultural biases. As expressed by P1, "\textit{Although I haven’t been there, I feel like when creating it, it refreshed my impression of it. Maybe the reality is not exactly like this, but my impression might be mixed with the images I generated and the images I’ve seen before, blending into a fusion of memories.}"

\section{Discussion}\label{sec:Discussion}

\subsection{The Impact of Generative AI on Personal Narrative Creation for Familiar Cultural Heritages}

Our findings revealed that participants held diverse perspectives and memories related to various heritage sites, including those deemed of relatively minor historical significance by heritage professionals. The collaboration with AI facilitated participants in visually presenting their experiences, memories, and understanding of familiar cultural heritage. These personal narratives may contribute to the dynamic historical information of cultural heritage and amplifying individual voices distinct from official institutions within the space ~\cite{tsenova_-authorised_2020}, offering fresh perspectives distinct from those presented in AHD ~\cite{bala_towards_2023,kambunga_participatory_2020}.

The creation of human-AI co-generated images involves an interactive collaboration between humans and AI. When participants attempted to express their understanding and emotions concerning cultural heritage, they often initially held only abstract concepts and fragmented memories. Generative AI can still generate complete images from vague keywords, allowing individuals to refine their ideas based on the output images ~\cite{gao_collabcoder_2023,liu_3dall-e_2023,zhang_generative_2023}. As demonstrated in section 4.2.1 of the results, participants' memories related to cultural heritage become increasingly vivid through continuous iterations ~\cite{claisse_crafting_2020} and the clarity of their themes progressively improved. Additionally, given that participants were not necessarily experts in narrative and visual communication, they often struggled to effectively convey their ideas through images ~\cite{zhao_involving_2023}. However, as demonstrated in Section 4.2.1, generative AI provided participants with feedback and creative insights, assisting them in refining their images ~\cite{weisz_better_2022,chiou_designing_2023}. This helped them find better ways to express their cultural narratives through images.

However, generative AI often generates results that do not align with participants' expectations during the generation process. Some mistakes may be attributed to the inherent limitations of generative AI, but participants, unaware of these limitations beforehand, may spend a significant amount of time without achieving the desired results. A potential solution to this issue is to provide participants with information of the reasons behind each generation mistake, as suggested by Jeung et al.~\cite{jeung_correct_2023}. This might help them to gain a better understanding of the capabilities and limitations of AI in generating diverse content. Our results indicated that participants, such as P17, who possessed this understanding, can promptly adjust their creative expectations and strategies. Another approach is to employ prompt engineering to test and optimize prompts, thereby enhancing the effectiveness and controllability of Stable Diffusion ~\cite{kim_alphadapr_2023,li_approach_2024}. This enables participants to input more precise prompts and achieve desired results. Additionally, reinforcement learning from human feedback can be used to fine-tune stable diffusion ~\cite{ouyang_training_2022,osman_leveraging_2023}, better aligning with people's preferences and requirements when creating cultural heritage narratives. These methods can serve as potential solutions to reduce the challenges faced by individuals unfamiliar with generative AI tools, thus increasing their accessibility to cultural narrative creation.

Additionally, it is important to note that in this workshop, we asked participants to engage in narrative through image creation. However, if they were required to supplement with other narrative elements, such as incorporating a written story, their creative results might differ. For instance, they might use new types of creative strategies, and their images might include more narrative elements instead of raw depictions.

\subsection{Generative AI Should Serve as a Suggestion Provider, Not a Controller}

The AI-generated outcomes contained uncertainties that introduced unexpected elements into the images. Sometimes these uncertainties provided valuable insights for participants' creations, such as the "first-person perspective scene" for P15 and the "red eyeball" for P18, but other times they were beyond the control of participants. As noted in Section 4.4.2 of the results, when creating images related to unfamiliar cultural heritage, participants lacking relevant knowledge were unable to identify mistakes related to the unfamiliar cultural heritage, which could result in AI leading the narrative process and leaving many erroneous uncertainties in the final narrative ~\cite{gao_collabcoder_2023}. However, when participants generated images related to familiar cultural heritage, as indicated in Section 4.4.1, they were able to continuously evaluate and select AI-generated results based on their own experiences and knowledge. This demonstrates that a foundational understanding of cultural heritage\cite{liu_digital_2023} can help mitigate mistakes in the images to some extent. 

Nonetheless, we must acknowledge that people's understanding of familiar cultural heritage can still contain subjective misconceptions or stereotypes. Therefore, to reduce inaccuracies introduced by both AI and individuals in the output images, it is beneficial to provide individuals with more objective references, such as actual photographs and official explanations of the cultural heritage. Additionally, collaborating with officials or staff in cultural heritage can help correct discrepancies between the images and objective facts.

\subsection{AI Exhibits Cultural Bias While Also Offering Cultural Inspiration}

To begin with, we need to contextualize this work within Chinese populations, as national identity may influence how narratives are generated. Chinese individuals generally hold a strong recognition of their culture ~\cite{xu2018cultivating} and want to maintain its independence and uniqueness within the global context. Consequently, when creating narratives, they often emphasized accurately conveying Chinese styles. For example, when P7 generated the street scene and P15 generated the clock tower, both emphasized Chinese-style features. They criticized that AI often falls short in capturing the unique features of Chinese cultural elements within the cultural heritage they aim to emphasize in their narratives. This includes AI-generated results with obvious Western rather than Chinese architectural characteristics and some that are filled with Western stereotypes of Chinese architecture. The inadequacy of AI's capabilities in this regard hinders the expression of these crucial and unique cultural aspects and exhibits stereotypes and inaccuracies in understanding identity descriptions related to specific cultures ~\cite{zhou_ethical_2023,bennett_its_2021}.

Surprisingly, however, we found that people's attitudes toward these cultural biases were not entirely negative. The bias introduced by AI also offered a fresh perspective, especially for participants familiar with the cultural heritage. Engaging with the AI-enabled process allowed participants to perceive their familiar cultural heritage from a new perspective and understand diverse interpretations of the same elements from other cultures ~\cite{jeon_rituals_2019}, such as P13 reflecting on the appearance features of deities from different cultures. This also implies that participants may have the opportunity to engage in critical reflection about their potential cultural biases towards other cultures through the biases present in the generated images related to their own culture. We hypothesize that being able to view cultural elements from the perspective of other cultures may enhance people's awareness of cultural diversity ~\cite{walsh_ai_2019}.

To address this challenge while enhancing its benefits, it is advisable to gather information about the unique cultural features of specific cultural heritages from communities or official sources. This ensures the curation of appropriate datasets for training more accurate models ~\cite{li_purposeful_2023}. Additionally, employing techniques such as prompt engineering and fine-tuning can enable the expression of different cultural characteristics more accurately through prompts. Continuously integrating individual narratives into the model can also help diversify the datasets, ensuring they reflect a broader range of voices, rather than being primarily shaped by influential or popular content driven by power.

\subsection{Limitations}
We have identified several limitations in this study. Firstly, we acknowledge that professional and cultural backgrounds may influence participants' use of narrative approaches, an aspect that we have not extensively explored. Future research can delve into the analysis of narrative approaches among individuals from diverse backgrounds, such as Eastern and Western backgrounds, to comprehend variations in AI narrative strategies across different cultural contexts. Additionally, our participants were primarily aged between 18-35 years and most had good proficiency with electronic devices. However, older individuals, who have rich narratives about cultural heritage, may not be able to learn how to use generative AI tools as quickly as the current participants. Future research could focus on how older age groups use AI tools to tell stories about cultural heritage, the strategies they employ, and the difficulties they encounter.

Secondly, the workshop format may not have fully allowed participants to openly discuss their familiar cultural heritage of interest. Utilizing alternative research methods, such as a series of workshops that provide more time for participants other functions of Stable Diffusion and refine their personal narratives, or incorporating gamified approaches to modify their collaborative interactions with generative AI, may yield different findings. 

Thirdly, our choice of Stable Diffusion as the AI generation tool may have influence on the creation process. Other generative AI tools, like Midjourney, could potentially have a different impact on participants' creation processes. For instance, Midjourney might be more user-friendly for novice users compared to Stable Diffusion, potentially altering their creation workflow. Additionally, using a multilingual dataset such as LAION to allow participants to input in their native language, as well as selecting a model tuned for landscapes to depict tangible cultural heritage, could also influence participants' creations.

Lastly, the unfamiliar cultural heritages provided to the participants were all from Hong Kong, which might have influenced their creation process due to their previous impressions of the region. Future research could consider diversifying the selection of unfamiliar cultural heritages to explore how this impacts participants' creation strategies.

\section{Conclusion}\label{sec:Conclusion}

We conducted workshops with 20 participants to explore their creative strategies using generative AI for creating narratives about familiar cultural heritages, as well as to evaluate both the strengths and limitations of generative AI in this context. We also examined how participants' familiarity with cultural sites influenced their creation by comparing the results of familiar and unfamiliar cultural heritages. The results showed three distinct narrative strategies adopted by participants. While generative AI proved beneficial in illuminating, amplifying, and reinterpreting personal narratives, its inherent cultural biases and uncertainties also posed challenges, particularly when participants lacked specific knowledge about the cultural heritages involved.

To address uncertainties and errors in AI-generated content, we recommend providing participants with detailed explanations for each generation mistake, utilizing prompt engineering, and applying reinforcement learning from human feedback to fine-tune Stable Diffusion. Moreover, to mitigate potential inaccuracies stemming from participants' inability to recognize AI-generated errors or subjective misconceptions, it is crucial to supplement their creative process with objective references such as authentic photographs and official descriptions of cultural heritages. Additionally, to enhance AI's ability to accurately capture and portray unique cultural elements, we propose the development of curated datasets tailored to specific cultural contexts for training more precise AI models.

Regarding the content of creation, workshop participants were asked to create representations of tangible cultural heritage. Future research may explore whether people achieve different results and creative strategies when using AI to describe intangible heritage-related themes that have a greater process component to the data material. Additionally, the participants' narrative expressions were limited to images. If we allow them to describe a story to assist in the narration of the images, their creation strategies might change.


\bibliographystyle{ACM-Reference-Format}
\bibliography{references}

 \newpage
\appendix
\label{sec:Appendix}
\section{Interview Questions}

\subsection{Pre-study interview question}

Where are you from?

Can you share something about [specific topic or place] or related to your personal experiences? / Can you imagine some Hong Kong CH that you have never visited before?

How familiar are you with [specific region or area]? Can you provide some information about [specific topic or place]? Also, do you know why this place is called [chosen CH]?

What were your initial thoughts when you first entered these places?

What emotional connections do you have with these places?

\subsection{Post-study interview question}

Why do you have these thoughts? Can you introduce your final work?

What are the differences between talking about or imagining CH and generating CH in SD?

Please evaluate the final generated work. Which image do you like the most, and why? How does it relate to your thoughts or imagination?

During the generation process, did you have any new feelings or thoughts about CH?

After the generation, do you have any other feelings or thoughts?

Why did you write this prompt? What kind of result do you hope to get from SD?

When you first tried to generate images of CH, what came to your mind initially?

What surprises did SD bring you during the generation process?

What difficulties did you encounter during the generation process? What do you wish SD could do?

Do you think the results matched your expectations?

Why did you choose these specific [topic] words? (Is it because these places have certain characteristics in your memory?)

What else do you think the entire generation process brought to you?

What are the main differences between these two generation processes?

\end{document}